\documentclass[aps,prb,preprint,groupedaddress]{revtex4-1}

\usepackage{graphicx,amsmath,amsfonts}
\usepackage{epstopdf}

\begin{document}

%\preprint{}

\title{Magnetic structure and ferroelectric activity in orthorhombic YMnO$_3$: relative roles
of magnetic symmetry breaking and atomic displacements}


\author{I. V. Solovyev}
\email{SOLOVYEV.Igor@nims.go.jp}
%\thanks{}
\affiliation{Computational Materials Science Unit,
National Institute for Materials Science, 1-2-1 Sengen, Tsukuba,
Ibaraki 305-0047, Japan}
\author{M. V. Valentyuk}
\author{V. V. Mazurenko}
\affiliation{
Department of Theoretical Physics and Applied Mathematics, Ural Federal University,
Mira str. 19, 620002 Ekaterinburg, Russia and Institute of Theoretical Physics, University of Hamburg,
Jungiusstrasse 9, 20355 Hamburg, Germany}


\date{\today}

\begin{abstract}
We discuss relative roles played by the magnetic inversion symmetry breaking and the
ferroelectric (FE) atomic displacements in the multiferroic state of YMnO$_3$.
For these purposes we derive a realistic low-energy model, using results of first-principles
electronic structure calculations and experimental parameters of the
crystal structure below and above the FE transition.
Then, we solve this model in the mean-field Hartree-Fock approximation.
We argue that the multiferroic state in YMnO$_3$ has a magnetic origin, and
the centrosymmetric $Pbnm$ structure is formally sufficient for explaining
main details of the noncentrosymmetric magnetic ground state.
The relativistic spin-orbit interaction lifts the degeneracy, caused by the frustration of isotropic
exchange interactions in the ${\bf ab}$ plane,
and stabilizes a twofold periodic noncollinear magnetic state, which is similar to the
E-state apart from the spin canting.
The noncentrosymmetric atomic displacements in the $P2_1nm$ phase reduce the spin canting,
but do not change the symmetry of the magnetic state.
The effect of the $P2_1nm$ distortion on the FE polarization $\Delta P_{\bf a}$, 
parallel to the orthorhombic ${\bf a}$ axis,
is twofold: (i) it gives rise to ionic contributions, associated with the oxygen and yttrium sites;
(ii) it affects the electronic polarization, mainly through the change of the spin canting.
The relatively small value of $\Delta P_{\bf a}$, observed in the experiment, is caused
by a partial cancelation of the electronic and ionic contributions, as well as different
contributions in the ionic part, which takes place for the experimental $P2_1nm$ structure.
The twofold periodic magnetic state competes with the fourfold periodic one and, even
in the displaced $P2_1nm$ phase, these two states continue to coexist in a narrow energy range.
Finally, we theoretically optimize the crystal structure.
For these purposes we employ the LSDA$+$$U$ approach and
assume the collinear E-type antiferromagnetic alignment. Then, use the obtained structural information again as
the input for the construction and solution of the low-energy model. We have found that
the agreement with the experimental data in this case is less satisfactory and $| \Delta P_{\bf a} |$
is largely overestimated. Although the magnetic structure
can be formally tuned by varying the Coulomb repulsion $U$ as a parameter,
apparently LSDA$+$$U$ fails to reproduce some fine details of the experimental structure,
and the cancelation of different contributions in $\Delta P_{\bf a}$ does not occur.
\end{abstract}

\pacs{77.55.Nv, 75.25.-j, 71.10.Fd, 71.15.Nc}
%\keywords{}

\maketitle

\section{\label{Intro} Introduction}

  Multiferroic materials, where the ferroelectric (FE) activity coexists with some long-range
antiferromagnetic (AFM)
order, have attracted a great deal of attention.\cite{MF_review}
A possibility to control magnetic properties by the electric field (and vice versa) makes
these materials promising for the creation of a new generation of electronic devices.
Moreover, there is a fundamental interest to the phenomenon of multiferroicity itself
and the microscopic origin of coupling between magnetic and electric degrees of freedom.

\begin{figure}
\begin{center}
\includegraphics[height=7cm]{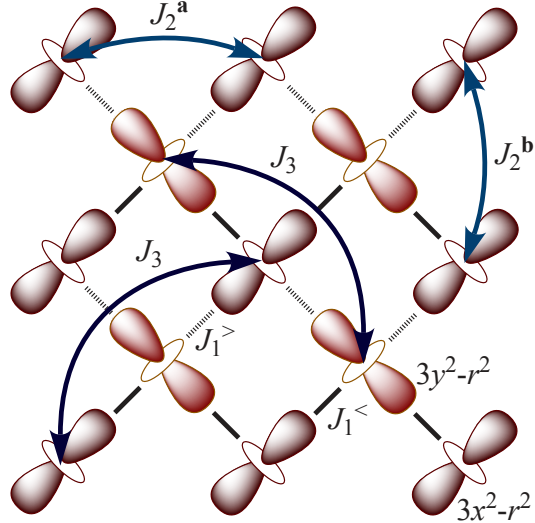}
\end{center}
\caption{\label{fig.intro}(Color online)
Schematic view on the
orbital ordering and magnetic interactions in the ${\bf ab}$ plane
of YMnO$_3$. $J_1^<$ and $J_1^>$ denote nearest-neighbor magnetic interactions,
associated with the compressed ($<$) and stretched ($>$) bonds
(shown by the filled and hatched lines, respectively) in the ferroelectric
$P2_1nm$ structure. $J_2^{\bf a}$ and $J_2^{\bf b}$ are the second-neighbor interactions along the
orthorhombic axes ${\bf a}$ and ${\bf b}$, respectively. $J_3$ is the strongest
third-neighbor interaction.}
\end{figure}

  The orthorhombic rare-earth manganites $R$MnO$_3$ are
one of the key compounds for understanding the basic concepts of such coupling.
There are two mechanisms of FE activity that are actively discussed
today.
One of them is the FE displacements in the
noncentrosymmetric
twofold periodic E-type
AFM ($\uparrow \uparrow \downarrow \downarrow$) state (Fig.~\ref{fig.intro}),\cite{SergienkoPRL}
which is caused by the exchange striction effects and leads to the alternation of the long and short
Mn-Mn bonds in the ${\bf ab}$ plane of $R$MnO$_3$.
This mechanism is nonrelativistic, and is believed to take place in the small-$R$ compounds, such as HoMnO$_3$ and YMnO$_3$.
Although the $\uparrow \uparrow \downarrow \downarrow$ arrangement itself is sufficient for the
magnetic inversion symmetry breaking, the FE displacements are believed to be
important for stabilizing the E-state.\cite{Picozzi}
Another mechanism 
is associated with the cycloidal spin-spiral alignment, which couples to the lattice
via the relativistic spin-orbit (SO) interaction.\cite{spiral_theories}
The characteristic compound, whose properties are typically regarded in the
context of this type of theories, is TbMnO$_3$.
It forms nearly fourfold periodic magnetic texture.\cite{TbMnO3_exp}

  In the recent publication (Ref.~\onlinecite{PRB11}), one of us has proposed that there is no
conceptual difference between twofold and fourfold periodic manganites,
and it is not quite right to consider separately 
the relativistic and nonrelativistic mechanisms of the FE activity.
Both types of magnetic structures are
formed due to the complex competition, involving three main ingredients: frustrated isotropic exchange interactions
of the nonrelativistic origin, single-ion anisotropy, and
Dzyaloshinskii-Moriya interactions.
Moreover, the magnetic structure, resulting from this competition, appears to be more complex, where
the collinear $\uparrow \uparrow \downarrow \downarrow$ E-type AFM alignment as well as
the cycloidal spin-spiral alignment can be regarded only as the first (and rather crude) approximation to the
true magnetic ground state. Particularly, it was argued that, even in the fourfold periodic TbMnO$_3$,
the FE activity
is caused by the \textit{deformation} of the homogeneous spin-spiral alignment (rather than the spin-spiral
alignment itself), and in this sense, the situation is similar to the twofold periodic HoMnO$_3$.\cite{PRB11}

  Regarding the twofold periodic magnetic ordering, realized in HoMnO$_3$ (which is similar to YMnO$_3$),
it was proposed that, although
the FE distortion apparently play an important role in stabilizing this magnetic state,
its symmetry is dictated by the competition of magnetic interactions in the
centrosymmetric $Pbnm$ phase.\cite{PRB11} This
competition is responsible for the magnetic inversion symmetry breaking, whereas the FE displacements should follow
this lowering of the magnetic symmetry.

  There are two important developments, which we would like to address in this work. First, the experimental
crystal structure,
realized below and above the FE transition (which coincides with the transition
to the E-state), has been reported recently
for YMnO$_3$.\cite{Okuyama} Particularly, it was shown that the FE transition at around 30~K
is accompanied by the lowering of symmetry from $Pbnm$ till $P2_1nm$. The experimental parameters of the crystal structure
and the atomic positions have been reported for two characteristic temperatures, 21~K and 50~K, correspondingly below and above
the transition.
Thus, following the idea of the previous work (Ref.~\onlinecite{PRB11}), we are able to clarify several aspects,
related to the formation of the twofold periodic magnetic texture
and the FE polarization in YMnO$_3$.
Particularly, what is the role of the $P2_1nm$ atomic displacements 
in the formation of the twofold periodic magnetic texture and the FE properties of YMnO$_3$?
Second, we have performed the structural optimization for the
E-type AFM state of YMnO$_3$, by using both
LSDA and LSDA$+$$U$ techniques (where, LSDA stands for the
local-spin-density approximation).
It allows us to address another important problem: all previous attempts to theoretically optimize
the crystal structure for the E-type AFM state systematically overestimated the value of the
FE polarization by an order of magnitude.\cite{Picozzi,Okuyama}
On the other hand, the theoretical calculations, which use the experimental $P2_1nm$
structure, seem to reproduce also the observed polarization value.\cite{Okuyama}
Therefore, it is very important to understand the origin of this problem
on a microscopic level.

  The rest of the paper is organized as follows. Sec.~\ref{Model} is devoted to the
analysis of the low-energy model, which was constructed by
using the experimental parameters of the crystal structure.
The method is briefly described in Sec.~\ref{Method}. Then, in Sec.~\ref{Sec:J} we
discuss the effect of the FE displacements in the $P2_1nm$ phase on
interatomic magnetic interactions. The summary of results without relativistic SO coupling is given in Sec.~\ref{Sec:TEP},
where we investigate the effect of the FE displacement on the total energy and
the electric polarization. In Sec.~\ref{Sec:GS} we consider the effect of the
relativistic SO interaction on the magnetic ground state, and in Sec.~\ref{Sec:P} we
discuss the value of the FE polarization.
Then, in Sec.~\ref{SOptimization}, we perform similar analysis, using results
of structural optimization for the collinear E-type AFM state, and
present a critical comparison with
the results, obtained by using the experimental crystal structure.
Finally, a brief summary of the work is given in Sec.~\ref{Summary}.

\section{\label{Model} Construction and solution of the low-energy model}

\subsection{\label{Method} Method}

  For the analysis of electronic and magnetic properties of YMnO$_3$, we employ
the same strategy as in previous publications, devoted to
multiferroic manganites.\cite{JPSJ,BiMnO3PRB10,PRB11}
First,
we construct an effective low-energy model (more specifically -- a multiorbital Hubbard model)
for the Mn $3d$ bands and derive parameters of this model from the first-principles
electronic structure calculations.
For these purposes, we use the experimental parameters of the $Pbnm$ and $P2_1nm$ structure.\cite{Okuyama}
Then, we use this model for the analysis of the magnetic ground state
and the FE polarization, which depends on the details of the magnetic and crystal structure of YMnO$_3$.

  The model itself has the following form:
\begin{equation}
\hat{\cal{H}}  =  \sum_{ij} \sum_{\alpha \beta}
t_{ij}^{\alpha \beta}\hat{c}^\dagger_{i\alpha}
\hat{c}^{\phantom{\dagger}}_{j\beta} +
  \frac{1}{2}
\sum_{i}  \sum_{\alpha \beta \gamma \delta} U_{\alpha \beta
\gamma \delta} \hat{c}^\dagger_{i\alpha} \hat{c}^\dagger_{i\gamma}
\hat{c}^{\phantom{\dagger}}_{i\beta}
\hat{c}^{\phantom{\dagger}}_{i\delta}.
\label{eqn.ManyBodyH}
\end{equation}
It is formulated in the basis of Wannier orbitals for the Mn $3d$ bands,
by starting from the electronic structure in the local-density approximation (LDA).\cite{review2008}
In these notations, each Greek symbol stand for the combination of spin
($s$$=$ $\uparrow$ or $\downarrow$) and orbital
($m$$=$ $xy$, $yz$, $3z^2$$-$$r^2$, $zx$, or $x^2$$-$$y^2$) variables.
The parameters of
the crystal field and the SO interaction
are included to the site-diagonal part of $t_{ij}^{\alpha \beta}$.
Off-diagonal matrix elements of
$t_{ij}^{\alpha \beta}$ with
respect to the site indices ($i$ and $j$) stand for the transfer integrals.
$U_{\alpha \beta \gamma \delta}$ are the screened Coulomb interactions.
The construction of the model (\ref{eqn.ManyBodyH}), definitions and details of
calculations of the model parameters
were explained in
the review article (Ref.~\onlinecite{review2008}).
The behavior of these parameters for the series of orthorhombic manganites
was discussed in previous publications (Refs.~\onlinecite{JPSJ,PRB11}).
The parameters, obtained for the $P2_1nm$ and $Pbnm$ phases of YMnO$_3$,
are collected in the supplemental materials.\cite{SM}

  After the construction, the model (\ref{eqn.ManyBodyH}) is solved
in the mean-field Hartree-Fock (HF) approximation:\cite{review2008}
$$
\left\{ \hat{t}({\bf k}) + \hat{\cal V} \right\} |C_{n {\bf k}} \rangle =
\varepsilon_{n {\bf k}} |C_{n {\bf k}} \rangle,
$$
where $\hat{t}({\bf k})$ is the Fourier image of $\hat{t}_{ij}$ and $\hat{\cal V}$
is the HF potential. The phases of $\hat{t}({\bf k})$ for the each pair of atoms
in the primitive cell are specified in the following way:
\begin{equation}
\hat{t}_{\boldsymbol{\tau}_i \boldsymbol{\tau}_j}({\bf k}) = \sum_{{\bf R}_j}
\hat{t}_{\boldsymbol{\tau}_i \boldsymbol{\tau}_j+{\bf R}_j}
e^{i{\bf k}\cdot(\boldsymbol{\tau}_j - \boldsymbol{\tau}_i + {\bf R}_j)},
\label{eqn:TPhases}
\end{equation}
where $(\boldsymbol{\tau}_j$$+$${\bf R}_j)$
is the position of the lattice site $j$ in terms of its coordinate $\boldsymbol{\tau}_j$
within the primitive cell and the lattice translation ${\bf R}_j$, and
$\hat{t}({\bf k})$ is the supermatrix:
$\hat{t}({\bf k}) \equiv \| \hat{t}_{\boldsymbol{\tau}_i \boldsymbol{\tau}_j}({\bf k}) \|$.

  The electric polarization is calculated by using the Berry-phase formalism:\cite{EPolarization}
\begin{equation}
\Delta P_a = \frac{1}{V}\sum_i Z_i \Delta \tau_{i,a} - \frac{1}{V} \frac{N_a}{N_1 N_2 N_3 |{\bf G}_a|}
\left[
\gamma_a(\infty) - \gamma_a(0)
\right].
\label{eqn:Pbasic}
\end{equation}
Here, the first term is the ionic contribution,
which is caused by the displacements ($\Delta \tau_{i,a}$) of the atomic charges ($Z_i$)
away from the centrosymmetric positions. It is calculated within the unit cell with the volume $V$.
The second term is the electronic contribution (in the following denoted as $\Delta {\bf P}^{\rm el}$),
which can be related to the difference of the Berry phases ($ \gamma $) in the process of
the adiabatic symmetry lowering.\cite{EPolarization}
In the case of improper multiferroics, where the inversion symmetry is broken
by some complex magnetic order, the `adiabatic symmetry lowering'
can be understood as the rotation of spins away from the centrosymmetric
ferromagnetic (FM)
configuration. For the ionic term, the reference point was constructed by
symmetrizing the atomic positions of the $P2_1nm$ structure, so that they
would transform to each other according to the symmetry operations of the
group $Pbnm$. In the process of such symmetrization,
the unit-cell volume was kept constant.

  The Berry phases can be evaluated on a discrete grid of the
${\bf k}$-points, generated by the $N_1$$\times$$N_2$$\times$$N_3$ divisions of the
reciprocal lattice vectors $\{ {\bf G}_a \}$ ($0 \leq s_a < N_a$)
$$
{\bf k}_{s_1,s_2,s_3} = \frac{s_1}{N_1}{\bf G}_1 + \frac{s_2}{N_2}{\bf G}_2 + \frac{s_3}{N_3}{\bf G}_3.
$$
This yields the construction of the type:\cite{EPolarization}
\begin{equation}
\gamma_1 = - \sum_{s_2 = 0}^{N_2-1} \sum_{s_3 = 0}^{N_3-1} {\rm Im} {\rm ln}
\mathbb{P}_{s_2 s_3},
\label{eqn:Bphase1}
\end{equation}
where similar expressions for $\gamma_2$ and $\gamma_3$ can be obtained by the cyclic permutation of the indices 1, 2, and 3.
Then, $\mathbb{P}_{s_2 s_3}$ can be related to the overlap matrices
$S = \| \langle C_{n {\bf k}} | C_{n' {\bf k}'} \rangle \|$ between
occupied eigenvectors $|C_{n {\bf k}} \rangle$ in two neighboring ${\bf k}$-points
(for instance, ${\bf k}$$=$${\bf k}_{s_1,s_2,s_3}$ and ${\bf k}'$$=$${\bf k}_{s_1+1,s_2,s_3}$
for $\gamma_1$, etc.). If $|C_{n {\bf k}} \rangle$ is periodic in the reciprocal space,
\begin{equation}
|C_{n {\bf k}+{\bf G}_a} \rangle = |C_{n {\bf k}} \rangle
\label{eqn:PeriodicGauge}
\end{equation}
(the so-called periodic gauge), $\mathbb{P}_{s_2 s_3}$ is given by the following expression:\cite{EPolarization}
\begin{equation}
\mathbb{P}_{s_2 s_3} =
\prod_{s_1 = 0}^{N_1-1} {\rm det} S({\bf k}_{s_1,s_2,s_3},{\bf k}_{s_1+1,s_2,s_3}),
\label{eqn:Bphase2}
\end{equation}
where, according to Eq. (\ref{eqn:PeriodicGauge}), $|C_{n {\bf k}} \rangle$ in the point ${\bf k}_{N_1,s_2,s_3}$
can be replaced by the one in the point ${\bf k}_{0,s_2,s_3}$. However,
if the kinetic part of the model Hamiltonian is given by Eq.~(\ref{eqn:TPhases}),
the eigenvectors are \textit{not} periodic. Therefore, we should modify
$\mathbb{P}_{s_2 s_3}$ by including the additional multiplier:
\begin{equation}
\mathbb{P}_{s_2 s_3} \rightarrow \mathbb{P}_{s_2 s_3} {\rm det} S({\bf k}_{N_1,s_2,s_3},{\bf k}_{0,s_2,s_3}).
\label{eqn:Bphase3}
\end{equation}
It describes the change of the phase at the zone boundary,
which appears when we reduce the
${\bf k}$-space 
integration to the first Brillouin zone.
Furthermore, since Eq.~(\ref{eqn:PeriodicGauge}) is no longer applicable,
the eigenvectors $|C_{n {\bf k}} \rangle$ in the point
${\bf k}_{N_1,s_2,s_3}$ should be obtained from independent diagonalization of the HF Hamiltonian.
Obviously that for the periodic gauge
$S({\bf k}_{N_1,s_2,s_3},{\bf k}_{0,s_2,s_3}) = 1$, and
this formulation is automatically reduced to the standard one (Ref.~\onlinecite{EPolarization}).
The correction (\ref{eqn:Bphase3}) was not considered in the previous works (Ref.~\onlinecite{BiMnO3PRB10,PRB11}).
However, it is found to be
small (basically, within the accuracy of the model analysis).
Nevertheless, in a number of cases, the extension (\ref{eqn:Bphase3})
can be important in order to reproduce the correct symmetry of the vector ${\bf P}$.

  Moreover, we have found a numerical error in the previous calculations of ${\bf P}$,
reported in Ref.~\onlinecite{BiMnO3PRB10,PRB11}: all numerical values reported
in these two papers should be additionally divided by
about $2.5$. This error was corrected in the present calculations.

  Finally, as will become clear below, the main
contribution to the electronic part of $\Delta {\bf P}$ 
is magnetic and is related to the formation of 
some complex magnetic texture. Therefore, we will also call it
`the magnetic part of $\Delta {\bf P}$'. Moreover, as we will see in Sec.~\ref{SOptimization},
the ionic polarization, parallel to the orthorhombic ${\bf a}$ axis, is associated with the oxygen and yttrium sites.
Thus, one can clearly separate the contributions of different atomic site:
yttrium and oxygen atoms
contribute to the ionic part of $\Delta P_{\bf a}$, while Mn atoms --
to the magnetic part of $\Delta P_{\bf a}$
(thereafter, we will denote by $\Delta P_{\bf a}$ the electric polarization
parallel to the orthorhombic ${\bf a}$ axis).

\subsection{\label{Sec:J} Behavior of interatomic magnetic interactions}

  In this section we briefly discuss the effect of the FE displacement
on the behavior of isotropic exchange interactions, without the relativistic SO coupling.

  Parameters of interatomic magnetic interactions, derived
using the perturbation theory expansion for infinitesimal
spin rotations near the FM state,\cite{review2008,Liechtenstein} are listed in Table~\ref{tab:J}.
\begin{table}[h!]
\caption{The values of interatomic magnetic interactions (in meV) in the ${\bf ab}$ plane of YMnO$_3$,
obtained for the experimental $P2_1nm$ and $Pbnm$ structures.
The notations of magnetic interactions are explained in Fig.~\protect\ref{fig.intro}.}
\label{tab:J}
\begin{ruledtabular}
\begin{tabular}{cccccc}
 crystal structure & $J_1^<$  & $J_1^>$  & $J_2^{\bf a}$ & $J_2^{\bf b}$ & $J_3$    \\
\hline
 $P2_1nm$          & $-$$5.8$ & $-$$3.7$ & $-$$0.2$      & $-$$1.2$      & $-$$2.2$ \\
 $Pbnm$            & $-$$5.0$ & $-$$5.0$ & $-$$0.2$      & $-$$1.2$      & $-$$2.3$ \\
\end{tabular}
\end{ruledtabular}
\end{table}
These parameters correspond to the following definition of the spin Hamiltonian:
${\cal H}_S = -\sum_{\langle ij \rangle} J_{ij} {\bf e}_i \cdot {\bf e}_j$, where
${\bf e}_i$ is the \textit{direction} of the spin at the site $i$, and the
summation runs over the \textit{inequivalent pairs} $\langle ij \rangle$.
In agreement with the previous finding,\cite{JPSJ,PRB11} the magnetic structure in the
${\bf ab}$ plane is mainly controlled by
$J_1$, $J_2^{\bf b}$, and $J_3$.

  The FE displacements in the $P2_1nm$ structure mainly affect the
nearest-neighbor (NN) interactions $J_1$.
Namely, since FE displacements create two types of the Mn-Mn bonds,
alternating along the ${\bf b}$ axis (see Fig.~\ref{fig.intro}),
the corresponding magnetic interactions become also inequivalent and
$J_1^<$, operating in the compressed bond, is generally different from
$J_1^>$, operating in the stretched one.
Then, we have obtained that these interactions satisfy the following inequality: $| J_1^< | > | J_1^> |$, meaning that
the compression of the Mn-Mn bonds strengthens the AFM coupling (and vice versa).
Such a behavior can be rationalized by considering the double exchange (DE) and
superexchange (SE) contributions to $J_1$. Of course, the concept of the DE physics is rather
unusual for insulators. We understand it in the following sense:
generally, in order to find
the energy gain, caused
virtual hoppings in an insulating state, one should employ the
SE theory. Then, in the ferromagnetically coupled bond,
this energy gain is proportional to
$t^2/\Delta$ (where $t$ is an effective transfer integral,
and $\Delta$ is the splitting between
occupied and empty levels with the same spin).
However,
in band insulators, $\Delta$ is created by the same hybridization effects as $t$.
Therefore, $\Delta \sim t$, and the energy gain is also proportional to $t$.
In this sense, the mechanism can be
called the `double exchange'.
The semiquantitative theory of DE and SE interactions
can be formulated in terms of the
$1/\Delta_{\rm ex}$ expansion for $J_1$,\cite{NJP08} where $\Delta_{\rm ex}$
is the intraatomic exchange splitting between the majority- and minority-spin states
(note that in manganites, $\Delta_{\rm ex}$ is
the largest parameter:
$\sim$$4.7$ eV in the case of YMnO$_3$).
Then, the DE describes the
zeroth-order effects with respect to $1/\Delta_{\rm ex}$ and the SE -- first-order effects.
Alternatively, one can say that the DE and SE are proportional to transfer integrals,
in the first
and second order, respectively.
The details can be found in Ref.~\onlinecite{NJP08}.

  Following this strategy, the DE interactions in the $P2_1nm$ structure can be evaluated as
$23.9$ meV and $23.2$ meV, for the compressed and stretched bonds, respectively
($23.8$ meV in the $Pbnm$ phase).
The SE interactions are of the order of $-$$30.7$ meV and $-$$27.9$ meV,
for the compressed and stretched bonds, respectively ($-$$29.6$ meV in the $Pbnm$ phase).
Thus, the change of $J_1$, as well as
difference between $J_1^<$ and $J_1^>$ in the FE state
is mainly controlled by the \textit{superexchange} processes.

  Similar tendency was found for the NN interactions between
adjacent ${\bf ab}$ planes,
where the compression of the bonds additionally stabilizes the AFM coupling.
For example, the interlayer coupling, operating in the compressed and stretched bonds of
the $P2_1nm$ phase, was found to be
$-$$7.3$ meV and $-$$6.8$ meV, respectively
(in comparison with $-$$7.1$ meV, obtained for the $Pbnm$ phase).

\subsection{\label{Sec:TEP} Ferroelectric displacements,
stability of the E-state,
and magnetic contribution to the electric polarization}

\begin{figure}
\begin{center}
\includegraphics[width=6.5cm]{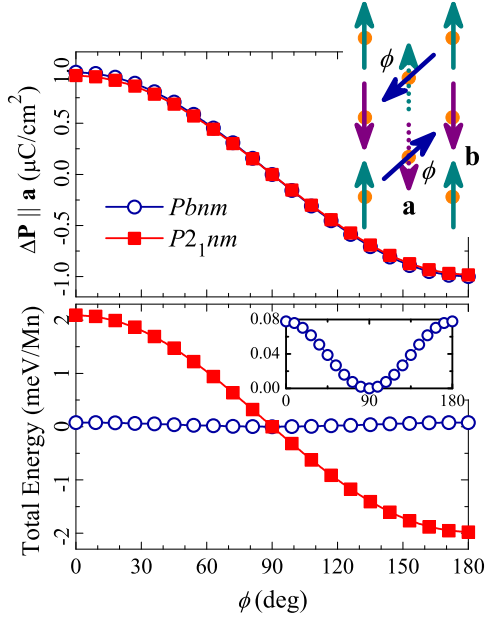}
\end{center}
\caption{\label{fig.2b}(Color online)
Results of Hartree-Fock calculations for the low-energy model,
constructed using experimental parameters for the
$P2_1nm$ and $Pbnm$ structure.
Upper panel shows the behavior of electronic polarization
for the general twofold periodic magnetic texture,
depending on the angle $\phi$
between Mn-spins in two magnetic sublattices.
The polarization is measured relative to the
spin-spiral state.
The magnetic structure
in the ${\bf ab}$ plane is explained in the inset.
Lower panel shows corresponding dependence of the total energy
on the angle $\phi$. The magnified behavior of the total energy
for the $Pbnm$ structure is shown in the inset (the notations and the scale of the
axes are the same as in the main figure).}
\end{figure}

  Fig.~\ref{fig.2b} shows results of HF calculations
for the general twofold periodic magnetic texture without SO coupling.
In this setting,
the spins in each magnetic sublattice are aligned antiferromagnetically along the ${\bf b}$ axis,
and two such sublattices in the ${\bf ab}$ plane are allowed to rotate relative to each other
(meanwhile, the magnetic coupling along the ${\bf c}$ axis was fixed to be AFM).
The relative direction of spins in two magnetic sublattices is specified by the angle $\phi$.
Then, $\phi = 90^\circ$ corresponds to the spin-spiral alignment,
while $\phi = 0$ and $180^\circ$ correspond to two AFM configurations of the E-type, which are degenerate
in the centrosymmetric $Pbnm$ structure, as was discussed in Ref.~\onlinecite{Picozzi}.
The details of calculations can be found in Ref.~\onlinecite{PRB11}.

  As expected, the FE displacements lift the degeneracy
of the E-states and stabilize only one of them (corresponding to $\phi = 180^\circ$
in our setting). In the ground-state configuration,
the AFM and FM coupling is set in
those Mn-Mn bonds, which are compressed and stretched, respectively.
This is consistent with the behavior of interatomic magnetic interactions,
discussed in Sec.~\ref{Sec:J}.
The total energy difference between stable ($\phi = 180^\circ$) and unstable ($\phi = 0$) E-states
is about $4.1$ meV/Mn. For the $Pbnm$ structure,
the total energy minimum corresponds to the spin spiral configuration, while the E-states are higher
in energy by about $0.08$ meV/Mn.\cite{PRB11} Thus, the FE displacements have a profound
effect on the total energy of orthorhombic YMnO$_3$.

   On the contrary, the magnetic state dependence of electronic polarization
is rather insensitive to whether the FE displacements are present or not.
In the centrosymmetric $Pbnm$ phase, the electronic polarization identically vanishes
for the spin-spiral state, $\Delta P_{\bf a}^{\rm el} = 0$, as expected without
the SO coupling.\cite{Picozzi}
In the $P2_1nm$ phase, $\Delta P_{\bf a}^{\rm el}(90^\circ)$ is finite. However, its
absolute value is more than one order of magnitude smaller than in the E-state.
Thus, to a good approximation, $\Delta P_{\bf a}^{\rm el}(90^\circ)$ can be used as
the reference point in the analysis of the magnetic-state dependence of
$\Delta P_{\bf a}^{\rm el}(\phi)$, both in the $Pbnm$ and $P2_1nm$ phase.
This magnetic-state dependence is shown in
Fig.~\ref{fig.2b}(a).
One can clearly see that the behavior of $\Delta P_{\bf a}^{\rm el}(\phi)$ is practically identical for the
$P2_1nm$ and $Pbnm$ structures.

  Thus, one can expect that the FE displacements in the $P2_1nm$ phase
may have only indirect effect on the magnetic part of the electric polarization:
they define the equilibrium value of $\phi$,
which controls $\Delta P_{\bf a}^{\rm el}$.
The direct contribution of the FE displacements to $\Delta P_{\bf a}^{\rm el}$ is small.
Nevertheless, they will contribute to the ionic polarization, which will be discussed below.

\subsection{\label{Sec:GS} Magnetic ground state}

\begin{figure}
\begin{center}
\includegraphics[height=4cm]{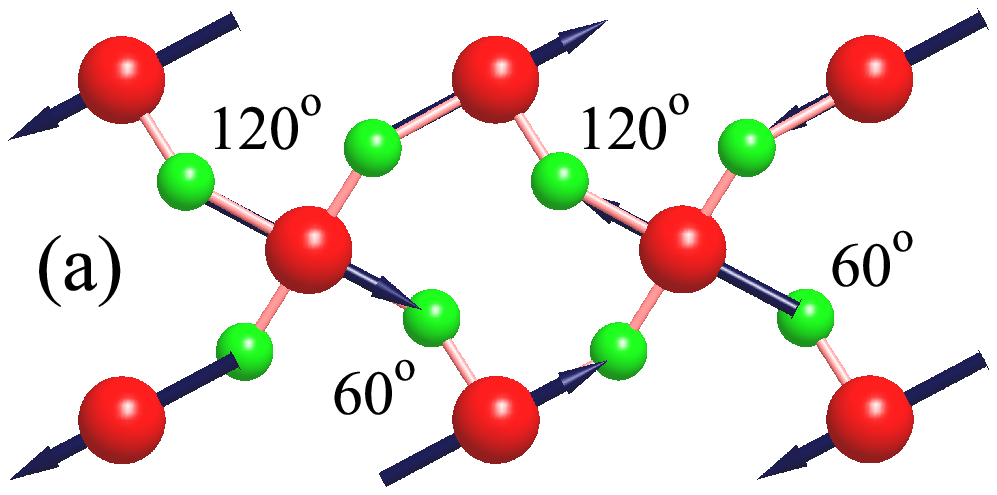}
\\
\includegraphics[height=4cm]{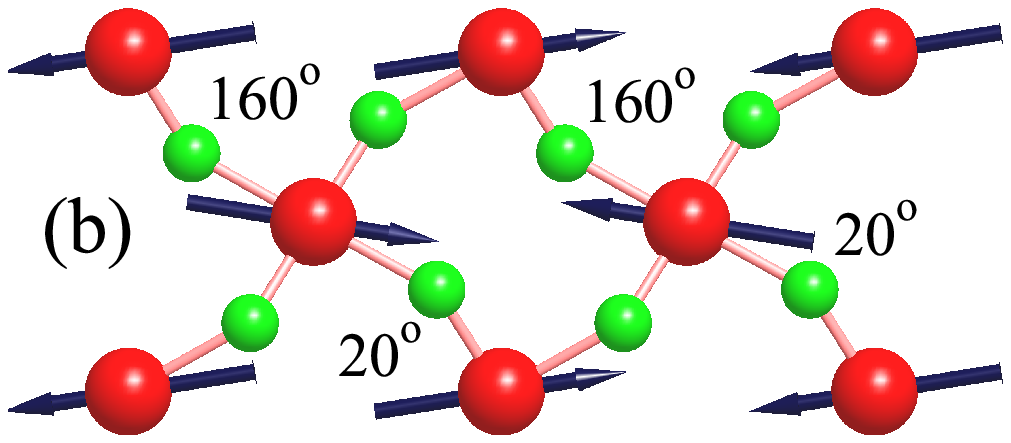}
\end{center}
\caption{\label{fig.SO2F}(Color online)
Distribution of spins in the ${\bf ab}$ plane of twofold periodic magnetic texture of YMnO$_3$,
as obtained for the
experimental $Pbnm$ (a) and $P2_1nm$ (b) structure.
The manganese atoms are indicated by the big red (dark) spheres and the oxygen atoms
are indicated by the small green (grey) spheres.
The numbers stand for the angles between spin magnetic moments in the
Mn-O-Mn bonds.
%For the better comparison, the experimental $P2_1nm$ structure has been rotated
%around the (horizontal) ${\bf b}$-axis by $180^\circ$.
}
\end{figure}

  In this section, we investigate the effect of the relativistic SO coupling on the magnetic 
ground state of YMnO$_3$.
We start from the collinear E-state, switch on the SO interaction, and further optimize the
magnetic texture, by solving the electronic low-energy model (\ref{eqn.ManyBodyH}) in the HF approximation
(the details can be found in Ref.~\onlinecite{PRB11}). The results of this optimization,
which was performed for the experimental $P2_1nm$ and $Pbnm$ structures,
are explained in Fig.~\ref{fig.SO2F}.
In both cases, the SO coupling leads to a noncollinear alignment.
Nevertheless, apart from some quantitative differences, two magnetic textures look rather similar.
Thus, the magnetic interactions, operating in the $Pbnm$ structure,
are already sufficient to
form the noncollinear magnetic texture, which breaks the inversion symmetry.
The relativistic SO interaction plays
a crucial role in this inversion symmetry breaking.
Indeed, let us consider the spin texture, which was obtained for the
$Pbnm$ structure (Fig.~\ref{fig.SO2F}a).
The spins in each magnetic sublattice are
ordered antiferromagnetically along the ${\bf b}$ axis. Such a magnetic order
is stabilized by
the magnetic interactions $J_2^{\bf b}$ and $J_3$ (see Fig.~\ref{fig.intro}).\cite{JPSJ} Then, in each magnetic
sublattice, the spins tend to align parallel to the longest Mn-O bonds and thus minimize the single-ion
anisotropy energy. In the combination with the AFM coupling along the ${\bf b}$ axis,
this will naturally lead to the alternation of angles between neighboring spins
($60^\circ$ and $120^\circ$, respectively) and the formation of the two inequivalent Mn-O-Mn bonds.
As soon as the bonds become inequivalent, the lattice will relax in order to adjust this change of the symmetry.
This qualitatively explains the origin of the FE displacements in the $P2_1nm$ structure.
Of course, the change of the crystal structure from $Pbnm$ to $P2_1nm$ will further affect
the magnetic texture, which becomes `more collinear' (the angles between neighboring spins
are $20^\circ$ and $160^\circ$, respectively).

  Another important issue is the competition between twofold and fourfold periodic magnetic textures.
The (nearly) fourfold periodic magnetic structure is realized, for example, in TbMnO$_3$.\cite{TbMnO3_exp}
In the previous work, one of us has argued that, in the centrosymmetric $Pbnm$ phase, the
fourfold periodic magnetic texture tends to have lower energy (both for TbMnO$_3$ and HoMnO$_3$,
although experimentally these two materials are known to form fourfold and
twofold periodic textures, respectively).\cite{PRB11} Therefore, it is reasonable to ask how this
tendency will be affected by the FE displacements in the $P2_1nm$ phase.
Intuitively, we expected that the FE displacements should stabilize
the twofold periodic magnetic texture, in agreement with the experiment.

\begin{figure}
\begin{center}
\includegraphics[height=4cm]{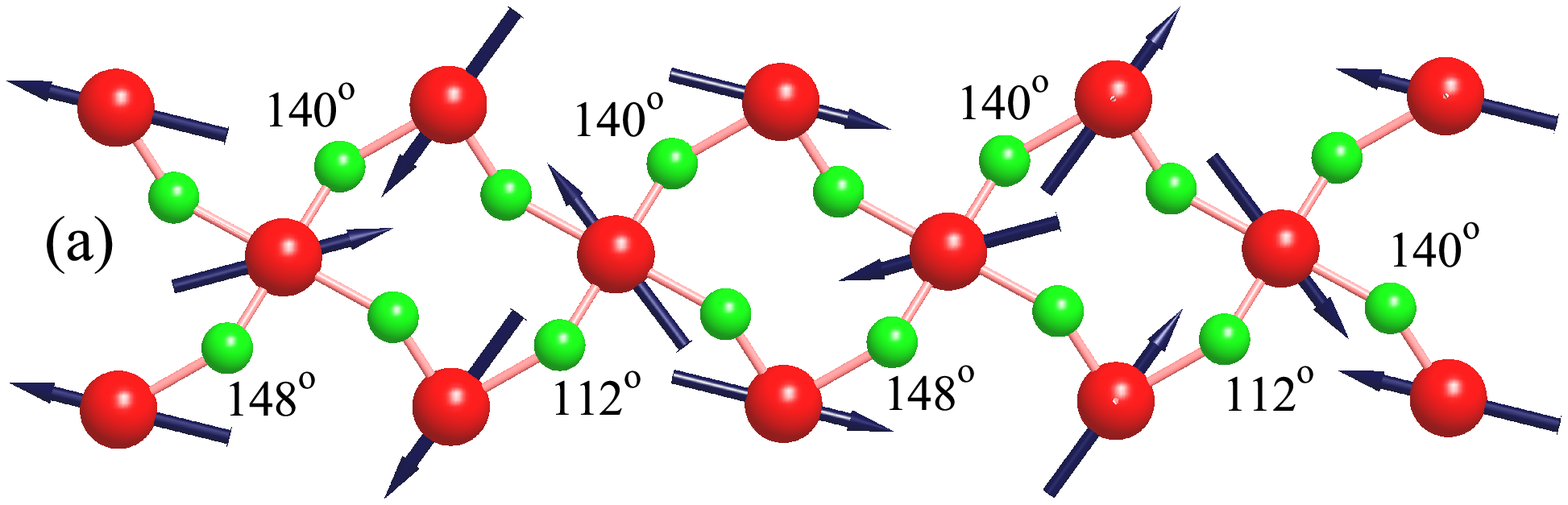}
\\
\includegraphics[height=4cm]{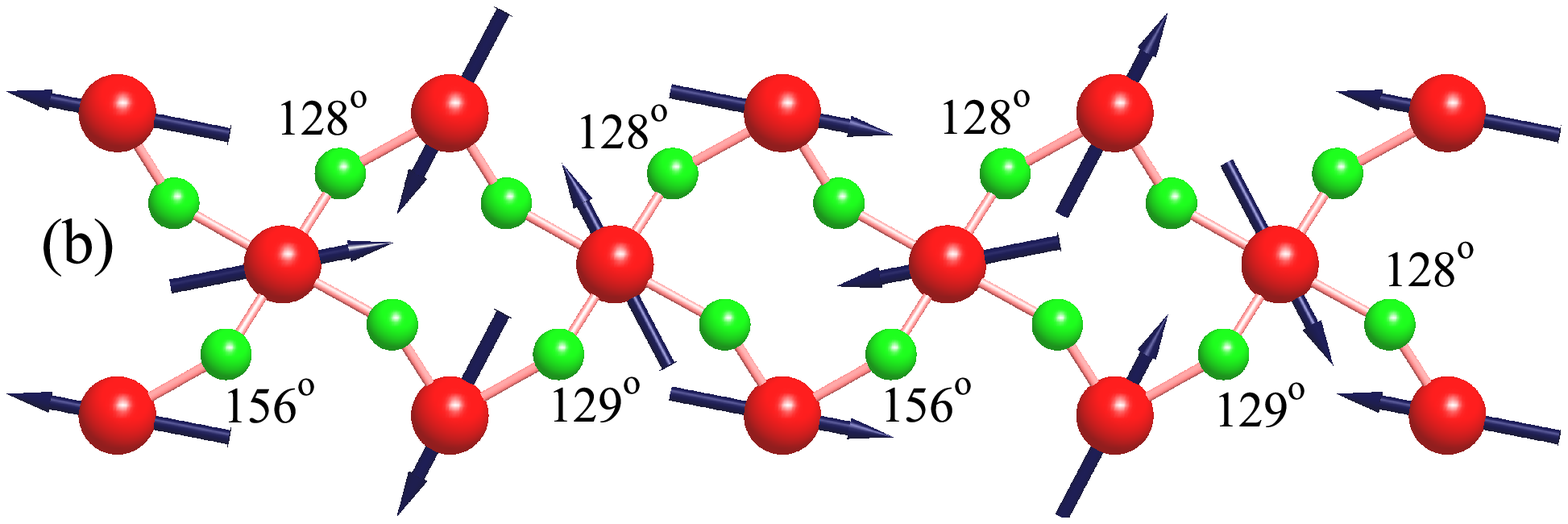}
\end{center}
\caption{\label{fig.SO4F}(Color online)
Distribution of spins in the ${\bf ab}$ plane of fourfold periodic magnetic texture of YMnO$_3$,
as obtained for the
experimental $Pbnm$ (a) and $P2_1nm$ (b) structure.
The manganese atoms are indicated by the big red (dark) spheres and the oxygen atoms
are indicated by the small green (grey) spheres.
The numbers stand for the angles between spin magnetic moments in the
Mn-O-Mn bonds.
%For the better comparison, the experimental $P2_1nm$ structure has been rotated
%around the (horizontal) ${\bf b}$-axis by $180^\circ$.
}
\end{figure}

  The fourfold periodic magnetic solution was obtained by starting from the
spin-spiral configuration (the so-called ${\bf ab}$ helix). Then, after switching on the SO coupling,
the spin spiral steadily equilibrates to the magnetic textures, which are
shown in Fig.~\ref{fig.SO4F}.
We would like to emphasize that there is a
substantial deviation from the spin-spiral alignment both for the $Pbnm$ and $P2_1nm$ phases.
The obtained textures are inhomogeneous in a sense that,
like in the twofold periodic case, they are also characterized by the alternation of angles between neighboring
spins in the ${\bf ab}$ plane.
For example, the magnetic textures in the $P2_1nm$ phase is characterized by three such angles:
$128^\circ$, $129^\circ$, and $156^\circ$ (for comparison, in the homogeneous
spin spiral, all angles are equal to $90^\circ$).
As was argued in Ref.~\onlinecite{PRB11}, this inhomogeneity plays a very important role
and is responsible for the
FE activity in the fourfold periodic magnanites, similar to the twofold
periodic ones.
Thus, we believe that there is no conceptual difference between
twofold and fourfold periodic
systems:
\begin{itemize}
\item[$\bullet$]
both superstructures have a magnetic
origin, where the relativistic SO interaction plays an important role;
\item[$\bullet$]
in both superstructures, the FE activity is caused by the
inhomogeneity of the magnetic texture or by its deviation from the conventional spin-spiral alignment.
\end{itemize}
In the $Pbnm$ phase of YMnO$_3$, the fourfold periodic texture has lower energy
than the twofold periodic one. This is consistent with results of the previous work.\cite{PRB11}
The total energy difference between these two states, obtained in the HF
approximation, is about $2.6$ meV/Mn.
Somewhat unexpectedly, we have found that, in
the $P2_1nm$ phase, the fourfold periodic texture still has lower energy, although
the total energy difference is reduced from $2.6$ till $1.0$ meV/Mn. Moreover, the distribution of
magnetic moments, obtained for these two crystallographic phases, is very similar (see Fig.~\ref{fig.SO4F}).
Thus, like in the twofold periodic case,
there is only a quantitative difference between 
fourfold periodic magnetic textures, realized in the $Pbnm$ and $P2_1nm$ phases.
This difference is also quite modest: for example, the angles between neighboring spins change only
within $10$-$20^\circ$.

  Probably, the precise analysis of the delicate balance between different magnetic states
is beyond abilities of our low-energy model, because the model itself does not
includes several ingredients, as was discussed in Ref.~\onlinecite{JPSJ}. Nevertheless, it clearly shows
that the FE displacements in the $P2_1nm$ phase work in the `right direction' and tend to additionally
stabilize the twofold periodic magnetic texture, in agreement with the experiment.
Moreover, the fact that the twofold and fourfold periodic magnetic states
are so close in energy may lead to the coexistence of these states, as was proposed recently
in some experimental works.\cite{Wadati}

\subsection{\label{Sec:P} Values of ferroelectric polarization}

  In this section we analyze different contributions to the FE polarization in the
$Pbnm$ and $P2_1nm$ phase of YMnO$_3$.
Obviously, the ionic contribution vanishes in the
centrosymmetric $Pbnm$ phase, but becomes finite in the
$P2_1nm$ one. Therefore, in the $P2_1nm$ structure, the definition of the
electronic and ionic terms
should be consistent with each other and also with the direction of
the FE displacement. Then, the first question is which values of the atomic
charges $\{ Z_i \}$ shall we use in our model analysis?
The first possibility is to take the ionic charges: i.e., $-$$2|e|$ for O and $+$$3|e|$ for Y and Mn.
In this case, the ionic contribution to $\Delta P_{\bf a}$ can be estimated as $0.55$ $\mu$C/cm$^2$.
The second possibility is to take $\{ Z_i \}$ from the
band structure calculations in the linear muffin-tin-orbital (LMTO) method,\cite{LMTO}
where each $Z_i$ is understood as the the positive nuclear charge plus negative electronic
charge within the atomic sphere $i$ (including empty spheres, which are used for a
better space filling). Then, the first term in Eq.~(\ref{eqn:Pbasic})
can be estimated as $-$$0.04$ $\mu$C/cm$^2$.
This simple analysis shows that the ionic part of $\Delta P_{\bf a}$ is sensitive
to the definition of $\{ Z_i \}$. Anyway, it seems to be smaller than the
electronic contribution to $\Delta P_{\bf a}$, caused by the noncentrosymmetric distribution of spins.

  The electronic polarization, obtained
for the twofold periodic magnetic texture in the $P2_1nm$ phase,
is about $-$$0.91$ $\mu$C/cm$^2$.
Then, the total polarization $\Delta P_{\bf a}$
can be estimated as $-$$0.95 \div -$$0.36$ $\mu$C/cm$^2$,
depending on the approximation, employed for the ionic part of $\Delta P_{\bf a}$.
These values are somewhat overestimated in comparison with the experimental data:
$| \Delta P_{\bf a} |$$=$ $0.24$ $\mu$C/cm$^2$ in the bulk (Ref.~\onlinecite{Okuyama})
and 0.80 $\mu$C/cm$^2$ in the thin films (Ref.~\onlinecite{Wadati}).
Nevertheless, the disagreement is quite modest: at least, we do not encounter the notorious
`one order of magnitude difference', which takes place if one uses the theoretically
optimized crystal structure instead of the experimental one.\cite{Picozzi,Okuyama}
We will come back to the analysis of this problem in Sec.~\ref{SOptimization}.
Moreover,
one should keep in mind two points:

(i) The electronic polarization in the twofold periodic texture is very sensitive to the
angle $\phi$ between spins in two magnetic sublattices.
Moreover, the closer the system to the spin-spiral alignment is,
the smaller the value of $| \Delta P_{\bf a}^{\rm el} |$ will be (Fig.~\ref{fig.2b}).
For example, in the $Pbnm$ phase, $\phi$ is closer to $90^\circ$ (Fig.~\ref{fig.SO2F}).
Therefore, the electronic
polarization is substantially smaller: $| \Delta P_{\bf a}^{\rm el} |$$=$ $0.55$ $\mu$C/cm$^2$.
Nevertheless, due to the FE displacements
in the $P2_1nm$ phase, the magnetic alignment becomes more collinear and the
value of the electronic polarization
increases. From this point of view, one reason why the experimental value of $| \Delta P_{\bf a} |$
is somewhat smaller than the theoretical one may be related to the fact that the experimental magnetic
texture itself is more noncollinear.
It would be interesting to check it experimentally, by measuring the deviation from the
collinear E-type AFM alignment.

(ii) Another mechanism, which might reconcile disagreement with the experimental data is the
coexistence of the two- and fourfold periodic magnetic textures, which was suggested recently
for the thin films of YMnO$_3$.\cite{Wadati}
According to our model analysis, the realization of the fourfold periodic magnetic texture in the
$P2_1nm$ phase (instead of the twofold periodic one) would lead to the drop of the electronic polarization from
$-$$0.91$ till $0.01$ $\mu$C/cm$^2$. Thus, the mixture of these two
magnetic states will probably decrease $| \Delta P_{\bf a} |$.

\section{\label{SOptimization}  Structural optimization: results and analysis}

  In this section we report results of structural optimization for the orthorhombic YMnO$_3$,
by assuming the collinear E-type AFM alignment, and discuss possible implications of the obtained structure
on the magnetic and FE properties. Namely, our strategy is the following.
First, we perform the structural optimization by using both LSDA and LSDA$+$$U$ functionals without the
relativistic SO coupling.
The main purpose of these calculations is to investigate the effect of the exchange striction,
caused by the collinear E-type AFM alignment. In this study, the value of $U$ was treated as
a parameter. Namely, we considered two such values: $U$$= 2.2$ eV -- an averaged value,
which is typically obtained for the low-energy model of orthorhombic manganites,\cite{JPSJ,SM}
and $U$$= 6.0$ eV, imitating the `large-$U$ limit'
(at least, the FE polarization in this regime is nearly saturated -- see Refs.~\onlinecite{Picozzi,Okuyama}).
As for the intraatomic exchange coupling, we take the atomic value
$J_{\rm H}$$=0.9$ eV.

  For the structural optimization, we use the projector augmented-wave method
implemented in the VASP code package.\cite{VASP}
The ${\bf k}$-space integration was performed on the
mesh of the $6$$\times$$6$$\times$$6$ points in the first Brillouin zone.
The plane wave cut-off energy was chosen to be 500 eV.
The optimization was performed by starting from the centrosymmetric experimental
$Pbnm$ structure.

  Then, we take the optimized crystal structure and again
construct the low-energy model for each set of the parameters, obtained for
different values of $U$. After that, we solve the
low-energy model and discuss the relative stability of the two- and fourfold periodic
textures as well as the different contributions to the FE polarization. Thus, we use LSDA$+$$U$
basically as a `machinery', which produces some crystal structure, and then we want to answer
the question how good is this crystal structure, by analyzing more delicate magnetic properties
and comparing them with the ones obtained for the experimental structure.

\begin{figure}
\begin{center}
\includegraphics[height=9cm]{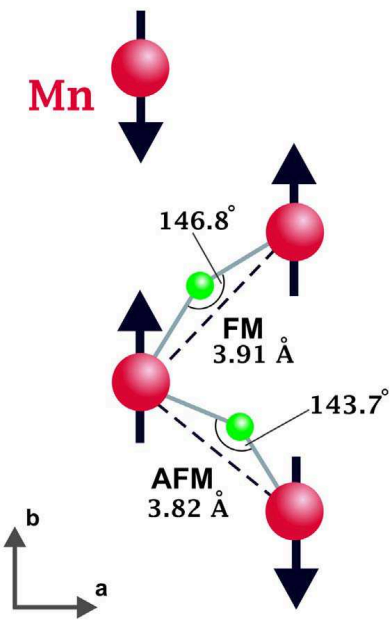}
\end{center}
\caption{\label{fig.Oscheme}(Color online)
Schematic representation of the Mn-O-Mn bonds in the E-type AFM structure.
The numerical values for the Mn-Mn distances and Mn-O-Mn angles
denote results of
structure optimization in LSDA$+$$U$ with $U$$= 2.2$ eV.
}
\end{figure}

  In Table~\ref{tab:structureU2}, we show
results of structural optimization with $U$$= 2.2$ eV. Similar results, obtained in
LSDA and LSDA$+$$U$ with $U$$= 6.0$ eV, are collected in Ref.~\onlinecite{SM}.
\begin{table}[h!]
\caption{Parameters of the $P2_1nm$ structure, obtained in LSDA$+$$U$ with $U$$= 2.2$ eV
for the E-type antiferromagnetic ordering. In this table, $x$, $y$, and $z$ are the fractional
coordinates, calculated in terms of the lattice parameters $a$$= 5.182$~\AA, $b$$= 5.740$~\AA, and $c$$= 7.251$~\AA, respectively.}
\label{tab:structureU2}
\begin{ruledtabular}
\begin{tabular}{lccc}
 site        & $x$     & $y$      & $z$              \\
\hline
Y1           & $0.478$ & $0.831$  & $1/2$            \\
Y2           & $0.520$ & $0.663$  & $0$              \\
Mn           & $0$     & $0.755$  & $0.251$          \\
O1           & $0.294$ & $0.569$  & $0.303$          \\
O2           & $0.887$ & $0.717$  & $1/2$            \\
O3           & $0.110$ & $0.791$  & $0$              \\
O4           & $0.698$ & $0.927$  & $0.196$          \\
\end{tabular}
\end{ruledtabular}
\end{table}
At first sight,
the internal coordinates, measured in units of the
orthorhombic lattice parameters, seem to agree with the experimental values reasonably well.\cite{Okuyama,remark1}
A more serious discrepancy occurs for the unit-cell volume,
which is underestimated by about 6~\% in LSDA and by 2~\% in LSDA$+$$U$ with $U$$= 6.0$ eV.
The values of Mn-Mn bondlengths and Mn-O-Mn angles in the ${\bf ab}$ plane,
obtained for different values of $U$,
are summarized in Table~\ref{tab:distances}. The notations are explained
in Fig.~\ref{fig.Oscheme}.
Although there is a correlation between
the theory and the experiment (for example, the Mn-Mn bondlengths and
Mn-O-Mn angles are generally larger for the FM bond, etc.), none of the calculations
reproduces the experimental values. LSDA overestimates the FE displacements:
both Mn-O-Mn angle and Mn-Mn bondlength are substantially smaller for the
the AFM bond in comparison with the FM one. This problem can be resolved by
including the Coulomb $U$, but only partially. For example,
the best estimate for
the AFM Mn-Mn bondlength was found for $U$$= 6.0$ eV. However, the FM Mn-Mn bondlength
practically does not depend on $U$. Both values are smaller than the experimental ones,
presumably because the unit-cell volume is also smaller. Moreover, the calculations
overestimate the difference between the Mn-O-Mn angles in the FM and AFM bonds,
even for $U$$= 6.0$ eV.
\begin{table}[h!]
\caption{Results of structural optimization for the E-type AFM structure in LSDA
and LSDA$+$$U$ with $U$$= 2.2$ and $6.0$ eV:
the Mn-Mn distances ($d$, in \AA) and Mn-O-Mn angles ($\angle$, in degrees),
as obtained for the FM and AFM bonds in the ${\bf ab}$ plane of YMnO$_3$,
in comparison with the experimental values (exp) reported in Ref.~\onlinecite{Okuyama}.
The geometry of the bonds is schematically explained in Fig.~\protect\ref{fig.Oscheme}.}
\label{tab:distances}
\begin{ruledtabular}
\begin{tabular}{lcccc}
 &  \multicolumn{2}{c}{AFM bond} & \multicolumn{2}{c}{FM bond} \\
                & $d$(Mn-Mn)  & $\angle$(Mn-O-Mn)  & $d$(Mn-Mn) & $\angle$(Mn-O-Mn) \\
\hline
LSDA            & $3.75$      & $143.8$            & $3.88$     & $148.5$   \\
$U$$=2.2$ eV    & $3.82$      & $143.7$            & $3.91$     & $146.8$   \\
$U$$=6.0$ eV    & $3.87$      & $143.9$            & $3.89$     & $145.2$   \\
exp             & $3.91$      & $143.8$            & $3.94$     & $144.1$   \\
\end{tabular}
\end{ruledtabular}
\end{table}

  Thus, it is interesting to ask how all these changes of the crystal structure
are reflected in the behavior of the FE polarization and the magnetic properties
of YMnO$_3$?

  The structural optimization in LSDA and LSDA$+$$U$
provides a reasonable starting point for the analysis of the noncollinear magnetic texture.
At least, the results are consistent with each other and can be interpreted in the
following way. Since LSDA overestimates the FE displacements, the obtained twofold
periodic magnetic texture is nearly collinear: the angles between NN spins in the ${\bf ab}$ plane
are 6$^\circ$ and 174$^\circ$ (Table~\ref{tab:mparam}), which are close to 0$^\circ$ and 180$^\circ$,
realized in the collinear E-state. The Coulomb $U$ decreases the FE displacements.
Therefore, the noncollinearity will systematically increase,
and $U$$= 6.0$ eV practically reproduces the values of $\phi$, obtained for the experimental $P2_1nm$ structure.
A similar behavior was found for the fourfold periodic magnetic texture: the LSDA
parameters of the crystal structure lead
to a nearly collinear magnetic alignment (the angles between NN spins are
11$^\circ$, 176$^\circ$, and 178$^\circ$, respectively -- Table~\ref{tab:mparam}), while the
noncollinearity again systematically increases with the increase of $U$.
\begin{table}[h!]
\caption{
Results of solution of four low-energy models, constructed for the
experimental $P2_1nm$ structure (exp) and theoretically optimized structures
in LSDA and LSDA$+$$U$ with $U$$= 2.2$ eV and $U$$= 6.0$ eV.
The structural optimization was performed by imposing the
collinear E-type antiferromagnetic alignment.
In this table, $\phi$ denotes the angles between neighboring Mn-spins in the ${\bf ab}$ plane
of twofold and fourfold periodic magnetic texture (denoted as
`2-fold' and `4-fold', respectively),
$\Delta P_{\bf a}^{\rm el}$ is the value of electronic polarization parallel to the orthorhombic ${\bf a}$ axis,
and $\Delta E$ is the total energy difference between two- and fourfold periodic
magnetic textures. For the experimental $P2_1nm$ structure, the distribution of the angles $\phi$ is explained in
Figs.~\ref{fig.SO2F} and \ref{fig.SO4F}.
}
\label{tab:mparam}
\begin{ruledtabular}
\begin{tabular}{ccccc}
 property                                            & exp           & LSDA         & $U$$= 2.2$ eV & $U$$= 6.0$ eV \\
\hline
 2-fold $\phi$ ($^\circ$)                            & 20, 160       & 6, 174       & 8, 172        & 23, 157       \\
 2-fold $\Delta P_{\bf a}^{\rm el}$ ($\mu$C/cm$^2$)  & $-$$0.91$     & $1.66$       & $0.86$        & $1.18$        \\
 4-fold $\phi$ ($^\circ$)                            & 128, 129, 156 & 11, 176, 178 & 31, 167, 171  & 68, 157, 157  \\
 4-fold $\Delta P_{\bf a}^{\rm el}$ ($\mu$C/cm$^2$)  & $0.01$        & $0.77$       & $0.36$        & $0.43$        \\
 $\Delta E$ (meV/Mn)                                 & $1.0$         & $-$$3.2$     & $-$$0.4$      & $0$           \\
\end{tabular}
\end{ruledtabular}
\end{table}

  However, the situation with the FE polarization appears to be more serious.
Let us start with the analysis of the ionic contribution, which was
computed by using ionic charges. Table~\ref{tab:partialI} shows
partial contributions to this polarization, associated with different atomic sites.
\begin{table}[h!]
\caption{Total and partial contributions to the ionic polarization
parallel to the orthorhombic ${\bf a}$ axis, originating from the
displacements of
4 yttrium atoms (4Y),
8 oxygen atoms in the ${\bf ab}$ plane (8O$_{\bf ab}$, denoted as O1 and O4 in Table~\protect\ref{tab:structureU2}), and
4 oxygen atoms between the planes (4O$_{\bf c}$, denoted as O2 and O3 in Table~\protect\ref{tab:structureU2}), which were obtained for the
experimental $P2_1nm$ structure (exp) and
theoretically optimized structures in LSDA and LSDA$+$$U$ with $U$$= 2.2$ eV and $U$$= 6.0$ eV.
All values
are in $\mu$C/cm$^2$. The structural optimization was performed by imposing the
collinear E-type antiferromagnetic alignment.}
\label{tab:partialI}
\begin{ruledtabular}
\begin{tabular}{lcccc}
 contribution  & exp                 & LSDA    & $U$$= 2.2$ eV       & $U$$= 6.0$ eV \\
\hline
 4Y            & $-$$0.35$           & $0.14$  & $-$$0.53$           & $0.08$        \\
 8O$_{\bf ab}$ & $\phantom{-}$$0.73$ & $4.34$  & $\phantom{-}$$2.61$ & $1.15$        \\
 4O$_{\bf c}$  & $\phantom{-}$$0.17$ & $0.09$  & $\phantom{-}$$0.33$ & $0.03$        \\
 total         & $\phantom{-}$$0.55$ & $4.57$  & $\phantom{-}$$2.41$ & $1.26$        \\
\end{tabular}
\end{ruledtabular}
\end{table}
Then, we can readily answer the question why the polarization, obtained for the experimental
$P2_1nm$ structure,
is relatively small (only $0.55$ $\mu$C/cm$^2$).
This is because there is a strong cancelation of contribution of the oxygen and yttrium sites
(note, that the contributions of manganese sites to the ionic polarization,
parallel to the orthorhombic ${\bf a}$ axis, vanish, because all
$\Delta \boldsymbol{\tau}_i$$\parallel$${\bf a} = 0$).
In the theoretical calculations, a similar situation was found only for $U$$= 2.2$ eV,
while in LSDA and for $U$$= 6.0$ eV all contributions are positive. Moreover,
the contributions of the oxygen atoms in the ${\bf ab}$ plane are strongly overestimated.
Therefore, the ionic polarization, obtained for the optimized theoretical structures,
is already larger than the experiment one.

  The disagreement becomes even more serious
when take into account the electronic polarization (Table~\ref{tab:mparam}),
which for all theoretical structures is aligned \textit{in the same direction} as the ionic one.
A similar situation was encountered in the previous calculations for HoMnO$_3$.\cite{Picozzi}
This behavior is clearly different
from the one obtained for the experimental $P2_1nm$ structure, where the ionic and
electronic terms were \textit{antiparallel}.

  The additional noncollinearity, obtained for the less distorted structures
(in the regime of large $U$), decreases $\Delta P_{\bf a}^{\rm el}$.
This tendency was found
both for the two- and fourfold periodic magnetic texture (Table~\ref{tab:mparam}).
Nevertheless, this behavior is nonmonotonous, and there is something special
in the theoretical `$U$$= 2.2$ eV' structure. First, only for this theoretical structure,
there is a partial cancelation of ionic contributions, associated with the
oxygen and yttrium atoms, similar to the experimental structure.
Then, this theoretical structure produces the minimal value of $\Delta P_{\bf a}^{\rm el}$, which
is also favorable from the viewpoint of comparison with the experimental polarization.
One the basis of these observations, one may speculate that $U$$= 2.2$ eV generates the best theoretical
crystal structure. However, the total polarization is largely overestimated,
and the agreement with the experimental data is still poor.

  The use of the LMTO charges, instead of the ionic ones,
does not resolve the problem: although the ionic polarization becomes somewhat smaller
(about $0.35$ $\mu$C/cm$^2$ for $U$$= 6.0$ eV), it has the same sign as
$\Delta P_{\bf a}^{\rm el}$. Thus the cancelation does not occur.
Moreover, in the ionic polarization, the contributions of the yttrium
and oxygen atoms also have the same sign and do not cancel each other.

  Apparently, the partial cancelation of the electronic and ionic polarizations, as well as the
contributions of the oxygen and yttrium sites to the
ionic part, explains relatively small value of $\Delta P_{\bf a}$,
which was observed in the experiment.
It also indicates at a serious problem,
existing
in the electronic structure calculations, which fail to reproduce some fine details of the
crystal structure, responsible for this cancelation.
One reason may be related with fundamental problems of LSDA or LSDA$+$$U$
(for instance, the problem of incorrect charge-transfer energy in LSDA$+$$U$, which cannot be
resolved simply by changing the value of $U$ -- Ref.~\onlinecite{PRB98}).
Another possibility is that the structural optimization should be performed for the
noncollinear magnetic texture, after including the SO coupling. However,
from the computational point of view,
this procedure is extremely difficult. For example, even without the structural relaxation,
the optimization of the magnetic structure in the low-energy model requires several thousands
of iterations. Obviously that simultaneous optimization of both crystal and magnetic structure
in the full scale electronic structure calculations will be much more
computationally demanding.

\section{\label{Summary} Summary and conclusions}

  We discussed the origin of the multiferroic state in the
twofold periodic orthorhombic manganites, by considering YMnO$_3$
as an example. Traditionally, the FE activity in these systems is attributed to the
exchange striction effects, associated with the collinear E-type AFM alignment.
In this picture, the FE displacements lift the frustration of isotropic exchange interactions
and stabilize the E-type AFM state, while the relativistic SO interaction is believed to be
irrelevant to this type of processes.

  Instead, we argued that the
E-state can be regarded as a combined effect of frustrated isotropic
interactions and the
single-ion anisotropy energy, acting in the ${\bf ab}$ plane of manganites.
The SO interaction plays a crucial role in this case: it lifts the degeneracy
and stabilizes a canted E-type AFM state. Formally speaking,
in order to form
the noncentrosymmetric
E-type AFM state, it is sufficient to have the centrosymmetric $Pbnm$ structure,
while the FE displacements
in the $P2_1nm$ structure
only
follow this lowering of the magnetic symmetry.
As far as the equilibrium magnetic structure is concerned, the use of the $P2_1nm$ structure instead of the
$Pbnm$ one yields only modest quantitative changes, while all main trends
can be obtained already in the centrosymmetric $Pbnm$ phase.

  In this analysis, we employed the realistic low-energy model,\cite{SM} derived from the first-principles electronic
structure calculations,\cite{review2008} and the experimental parameters of the crystal structure.\cite{Okuyama}

  Then, we addressed an important issue of the `order of magnitude difference' between
the theoretical and experimental values of the FE polarization $\Delta P_{\bf a}$, which is typically observed for the
twofold periodic systems.\cite{Picozzi} We have found that, if one uses the experimental $P2_1nm$
structure, one can get a reasonable agreement also with the experimental value for the FE polarization
(similar conclusion was drawn recently in Ref.~\onlinecite{Okuyama}).
The reason for it is the partial cancelation of different contributions in $\Delta P_{\bf a}$,
which takes place for the experimental $P2_1nm$ structure. Particularly, there is a cancelation of
electronic and ionic parts, as well as the contributions of the oxygen and yttrium atoms
to the ionic polarization.

  In the last part of this work, we attempted to explore the same questions on a purely theoretical level,
and derived parameters of the noncentrosymmetric $P2_1nm$ phase
from the first-principles calculations. For these purposes we employed the LSDA and LSDA$+$$U$ techniques
without the relativistic SO coupling, and performed the structural optimization by assuming the
collinear E-type AFM alignment. After that, we used again the obtained structural information for the
construction and solution of the low-energy model.
We have found that, although the obtained theoretical structure stabilizes the
canted E-type AFM state, it fails
to reproduce the experimental values of the FE polarization. Thus, it seems that whenever we use the
theoretical parameters of the crystal structure, we face the `order of magnitude difference'
problem for the FE polarization. Namely, the cancelation, which took place for the
experimental $P2_1nm$ structure, does not occur for the theoretical one.
Intuitively, this result is in line with our main idea:
if the SO interactions plays a crucial role in stabilizing the (canted) twofold periodic E-state, it cannot
be ignored in the process of the structural optimization. On the other hand, there may be more
fundamental problems, related to the LSDA and LSDA$+$$U$ functionals.
We hope that our analysis will stimulate further theoretical and experimental works aiming to resolve
this interesting and important issue.

  \textit{Acknowledgements}.
The work of MVV and VVM is supported by the grant program of President of Russian Federation MK-406.2011.2, the scientific program ``Development of scientific potential of Universities'', RFFI 12-02-90810, the grant of the Ministry of education and science of RussiaÊ N 12.740.11.0026 and partially supported by Deutsche Forschungsgemeinschaft (SFB 925).


\begin{thebibliography}{99}
\bibitem{remark1}
Note that in the
$P2_1nm$ structure, the reference point for the $x$-coordinate
can be chosen arbitrarily. In our case, it is more convenient to chose $x = 0$
for the $4b$ site, occupied by Mn.
\end{thebibliography}
\end{document}